\documentclass[nohyper,12pt,letterpaper]{JHEP3}
\usepackage{epsfig}
\usepackage[latin1]{inputenc}
\usepackage{bbm,amsfonts}
\usepackage{graphicx}
\usepackage{amssymb,amsmath}

\author{Marco S. Bianchi$^\ast$,
  Gaston Giribet$^\hash$,
  Matias Leoni$^{\hash}$,  
  and Silvia Penati$\dag$\\\\
  $^\ast$Institut f\"ur Physik,
Humboldt-Universit\"at zu Berlin,
Newtonstra{\ss}e 15, 12489 Berlin, Germany \\\\
  $^\hash$Physics Department, FCEyN-UBA \& IFIBA-CONICET\
Ciudad Universitaria, Pabell\'on I, 1428, Buenos Aires, Argentina \\\\
  $^\dag$Dipartimento di Fisica, Universit\`a degli studi di Milano--Bicocca and
  INFN, Sezione di Milano--Bicocca, Piazza della Scienza 3, I-20126 Milano, Italy \\\\
  \qquad\\\\
  E-mail: \email{ marco.bianchi@physik.hu-berlin.de, gaston@df.uba.ar, 
  leoni@df.uba.ar,
    silvia.penati@mib.infn.it}
}

\abstract{ We revisit the computation of the two--loop light--like tetragonal Wilson loop for three dimensional pure 
Chern--Simons and $\mathcal{N}=6$ Chern--Simons--matter theory, within dimensional regularization with dimensional 
reduction scheme. Our examination shows that, contrary to prior belief, the result respects maximal transcendentality 
as is the case of the four--point scattering amplitude of the theory. Remarkably, the corrected result matches exactly 
the scattering amplitude both in the divergent and in the finite parts, constants included. }

\preprint{April 2013\\ HU-EP-13/21}

\title{Light--like Wilson loops in ABJM and maximal transcendentality.}


\csname @addtoreset\endcsname{equation}{section}

\def\bseq{\begin{subequation}}  
\def\eseq{\end{subequation}}
\def\bsea{\begin{subeqnarray}}  
\def\esea{\end{subeqnarray}}

\newcommand{\beq}{\begin{equation}}
\newcommand{\bea}{\begin{eqnarray}}
\newcommand{\eea}{\end{eqnarray}}
\newcommand{\eeq}{\end{equation}}

\newcommand {\non}{\nonumber}

\renewcommand{\b}{\beta}

\newcommand{\D}{\Delta}
\newcommand{\e}{\epsilon}

\def\Mb{\kern 2pt\mathchoice
        {
         \vbox{\hrule width10pt height 0.4pt depth 0pt
         \kern 1.2pt\hbox{\kern -2pt$\displaystyle M$}}}
        {
         \vbox{\hrule width10pt height 0.4pt depth 0pt
         \kern 1.2pt\hbox{\kern -2pt$\textstyle M$}}}
        {
\vbox{\hrule width6pt height 0.4pt depth 0pt
         \kern 1.0pt\hbox{\kern -2pt$\scriptstyle M$}}}
        {
         \vbox{\hrule width5pt height 0.4pt depth 0pt
         \kern 0.8pt\hbox{\kern -2pt$\scriptscriptstyle M$}}}}

\def\Sb{\kern 2pt\mathchoice
        {
         \vbox{\hrule width6pt height 0.4pt depth 0pt
         \kern 1.2pt\hbox{\kern -2pt$\displaystyle S$}}}
        {
         \vbox{\hrule width6pt height 0.4pt depth 0pt
         \kern 1.2pt\hbox{\kern -2pt$\textstyle S$}}}
        {
         \vbox{\hrule width3.5pt height 0.4pt depth 0pt
         \kern 1.0pt\hbox{\kern -2pt$\scriptstyle S$}}}
        {
         \vbox{\hrule width3pt height 0.4pt depth 0pt
         \kern 0.8pt\hbox{\kern -2pt$\scriptscriptstyle S$}}}}

\def\Rb{\kern 2pt\mathchoice
        {
         \vbox{\hrule width5.5pt height 0.4pt depth 0pt
         \kern 1.2pt\hbox{\kern -2.5pt$\displaystyle R$}}}
        {
         \vbox{\hrule width5.5pt height 0.4pt depth 0pt
         \kern 1.2pt\hbox{\kern -2.5pt$\textstyle R$}}}
        {
         \vbox{\hrule width3.5pt height 0.4pt depth 0pt
         \kern 1.0pt\hbox{\kern -2.2pt$\scriptstyle R$}}}
        {
         \vbox{\hrule width3pt height 0.4pt depth 0pt
         \kern 0.8pt\hbox{\kern -2.2pt$\scriptscriptstyle R$}}}}

  \def\pp{{\mathchoice
      {
          \kern 1pt%
          \raise 1pt
          \vbox{\hrule width5pt height0.4pt depth0pt
            \kern -2pt
            \hbox{\kern 2.3pt
              \vrule width0.4pt height6pt depth0pt
              }
            \kern -2pt
            \hrule width5pt height0.4pt depth0pt}%
            \kern 1pt
       }
        {
          \kern 1pt%
          \raise 1pt
          \vbox{\hrule width4.3pt height0.4pt depth0pt
            \kern -1.8pt
            \hbox{\kern 1.95pt
              \vrule width0.4pt height5.4pt depth0pt
              }
            \kern -1.8pt
            \hrule width4.3pt height0.4pt depth0pt}%
            \kern 1pt
        }
        {
          \kern 0.5pt%
          \raise 1pt
          \vbox{\hrule width4.0pt height0.3pt depth0pt
            \kern -1.9pt  
            \hbox{\kern 1.85pt
              \vrule width0.3pt height5.7pt depth0pt
              }
            \kern -1.9pt
            \hrule width4.0pt height0.3pt depth0pt}%
            \kern 0.5pt
        }
        {
          \kern 0.5pt%
          \raise 1pt
          \vbox{\hrule width3.6pt height0.3pt depth0pt
            \kern -1.5pt
            \hbox{\kern 1.65pt
              \vrule width0.3pt height4.5pt depth0pt
              }
            \kern -1.5pt
            \hrule width3.6pt height0.3pt depth0pt}%
            \kern 0.5pt
        }
    }}

  \def\mm{{\mathchoice
               {
                 \kern 1pt
           \raise 1pt    \vbox{\hrule width5pt height0.4pt depth0pt
                  \kern 2pt
                  \hrule width5pt height0.4pt depth0pt}
                 \kern 1pt}
               {
                \kern 1pt
           \raise 1pt \vbox{\hrule width4.3pt height0.4pt depth0pt
                  \kern 1.8pt
                  \hrule width4.3pt height0.4pt depth0pt}
                 \kern 1pt}
               {
                \kern 0.5pt
           \raise 1pt
                \vbox{\hrule width4.0pt height0.3pt depth0pt
                  \kern 1.9pt
                  \hrule width4.0pt height0.3pt depth0pt}
                \kern 1pt}
               {
               \kern 0.5pt
         \raise 1pt  \vbox{\hrule width3.6pt height0.3pt depth0pt
                  \kern 1.5pt
                  \hrule width3.6pt height0.3pt depth0pt}
               \kern 0.5pt}
               }}

\def\pd{{\kern0.5pt
           + \kern-5.05pt \raise5.8pt\hbox{$\textstyle.$}\kern
0.5pt}}

\def\pmd{{\kern0.5pt
          \pm \kern-5.05pt
\raise6.3pt\hbox{$\textstyle.$}\kern1.5pt}}

\def\md{{\mathchoice
   {
      {{\kern 1pt - \kern-6.2pt \raise5pt\hbox{$\textstyle.$}\kern
1pt}}}
    {
      {{\kern 1pt - \kern-6.2pt \raise5pt\hbox{$\textstyle.$}\kern
1pt}}}
    {
      {\kern0.5pt - \kern-5.05pt
\raise3.4pt\hbox{$\textstyle.$}\kern0.5pt}}
    {
      {\kern0.5pt - \kern-5.05pt
\raise3.4pt\hbox{$\textstyle.$}\kern0.5pt}}}}

\def\beq{\begin{equation}}
\def\eeq{\end{equation}}
\def\bea{\begin{eqnarray}}
\def\eea{\end{eqnarray}}

\def\b{\beta}

\def\e{\epsilon}

\def\D{\Delta}

\begin{document}

\section{Preliminaries}

Wilson loops are important non--local operators in any gauge theory. In three dimensions they play a central role in 
that their expectation values and correlation functions constitute the main observables of Chern--Simons theories.

In this note we will focus on Wilson loops evaluated on a light--like polygonal contour, which display remarkable 
properties. In particular, they develop UV divergences due to the cusps in the contour, which get strengthened by the 
on--shell conditions on the edges. These divergences are controlled by a rather universal and ubiquitous quantity, the 
cusp anomalous dimension, which also governs the leading IR singularities of scattering amplitudes of massless 
particles and the anomalous dimensions of high--spin operators.

In ${\cal N}=4$ SYM, light--like Wilson loops exhibit even more interesting features, as it has been observed that, 
loop by loop, not only their UV divergent part maps to the IR divergent part of scattering amplitudes, but also their 
finite non--constant contributions match \cite{Drummond:2007aua}--\cite{Anastasiou:2009kna}. This equivalence is achieved 
by writing particle momenta in scattering amplitudes in terms of dual variables $p_i=x_{i+1}-x_i$, which are then 
identified with the ordinary space--time coordinates of Wilson loops. Under this transformation ordinary conformal invariance of Wilson loops is mapped into the so--called dual conformal invariance of scattering amplitudes. 

The Wilson loops/scattering amplitudes duality has a profound meaning at strong coupling. In this regime, scattering in ${\cal 
N}=4$ SYM was studied via the AdS/CFT correspondence, where dual coordinates emerge as a T--duality transformation and 
the computation of the amplitude is mapped to that of a Wilson loop with a light--like polygonal shape \cite{AM}. Moreover, the duality has a deep explanation in terms of  the invariance of the dual superstring model under a combination of bosonic and fermionic T--dualities \cite{BM,Beisert:2008iq}. 

All known perturbative results for scattering amplitudes and light--like Wilson loops in ${\cal 
N}=4$ SYM respect the {\em maximal transcendentality principle} \cite{Kotikov:2001sc}, first formulated for the anomalous dimensions of twist--2 operators \cite{Kotikov:2001sc}--\cite{Kotikov:2012ac}. 
In the case of scattering amplitudes and Wilson loops computed using the dimensional reduction scheme, this principle states that assigning transcendentality $(-1)$ to the dimensional regularization parameter $\epsilon$, the $l$--loop correction to such observables exhibits uniform transcendentality $2l$. 
More generally, the principle has been proved to be satisfied by other quantities in ${\cal N}=4$ SYM, like for instance the Sudakov form 
factor \cite{Gehrmann:2011xn}. This leads to speculate that the principle might be related to intrinsic properties of the master integrals that appear in this theory \cite{Kotikov:2012ac}. In particular, for planar scattering amplitudes maximal transcendentality seems to be a property of dual conformally invariant integrals appearing in their loop computation.

It is certainly interesting to investigate whether such a remarkable property is shared by other theories in different dimensions 
and/or with a different amount of supersymmetry. In this note we will address this question for pure Chern--Simons 
theories in three dimensions and for the ${\cal N}=6$ Chern--Simons--matter theory (ABJM) introduced in \cite{ABJM}. 

Known results for tree amplitudes in ABJM  \cite{BLM,HL} signal the presence of dual superconformal \cite{Drummond:2006rz}--\cite{Brandhuber:2008pf} and Yangian \cite{DHP} symmetry. By using a three--dimensional version of BCFW recursion relations \cite{Gang:2010gy} these properties could be extended to planar loop integrands. In fact, the two--loop four point amplitude \cite{CH, BLMPS1} and one--loop amplitudes \cite{Bargheer:2012cp, Bianchi:2012cq} can be written as linear combinations of dual conformally invariant integrals. Notably, the corresponding results exhibit maximal transcendentality. Perturbative results on form factors in ABJM are going to appear \cite{Travaglini}, exhibiting this property too.
On the other hand, ${\cal N}=8$ SYM in three dimensions violates the maximal transcendentality principle. In fact, its two--loop four--point scattering amplitude contains terms of non--uniform transcendentality \cite{Bianchi:2012ez}. This is probably connected to the fact that the integrals in this theory are just dual conformally covariant but not invariant \cite{Lipstein}. 

Light--like Wilson loops in three--dimensional Chern--Simons theories with and without matter have been computed in 
perturbation theory up to two loops. In \cite{BLMPRS} their expectation value has been found to vanish at one loop for 
polygonal contours with any number of edges. In \cite{HPW} the authors have derived an expression for the two loop 
correction to the four cusps Wilson loop in pure Chern--Simons and in ABJM theory, which has been later extended to all 
points \cite{Wiegandt}.

In three dimensions, strong arguments in support of dual superconformal symmetry and Wilson 
loops/scattering amplitudes are still lacking. In fact, at strong coupling it is not yet
 clear how to implement the fermionic T--duality invariance of the ABJM dual string background 
\cite{ADO}--\cite{Colgain:2012ca}. At weak coupling, despite a number of results on 
higher--point amplitudes has been found in \cite{Bargheer:2012cp, Bianchi:2012cq, Brandhuber:2012un, yutin}, it is not clear yet how the duality might work, since for more than four external particles the amplitudes cease to be MHV \footnote{In three dimensions there is no notion of helicity. However, the name MHV (Maximally-Helicity-violating) is commonly used in analogy to 
${\cal N}=4$ SYM amplitudes.}. 

The four cusps Wilson loop in ABJM is nonetheless special since it has been discovered to match the non--constant part of the 
two--loop four--point amplitude \cite{CH, BLMPS1, BLMPS2, BLP}. This is a hint that a Wilson loop/scattering amplitude duality might work even for ABJM.  

However, a puzzle already highlighted in \cite{HPW} arises for what concerns maximal transcendentality: 
The Wilson loop result fails to be maximally transcendental due to a residual $(\log 2)$ term 
appearing in the constant, which cannot be reabsorbed into a redefinition of the regularization scale. On the contrary, 
the four--point amplitude seems to enjoy dual conformal symmetry and respects the maximal transcendentality principle.
Therefore, it would be quite inexplicable if the dual Wilson loop were not maximally transcendental.

It is the purpose of this note to tackle and solve this puzzle. We will argue that the problem resides in the 
regularization of a divergent integral associated to a gauge three--vertex diagram. After regularizing it by 
dimensional regularization in dimensional reduction scheme (DRED) and treating properly the contractions between 
three--dimensional and $d$--dimensional objects, we find that maximal transcendentality is restored, opening the possibility 
that this property can hold for pure Chern--Simons and ABJM, as well.

\section{Regularization}

We will be primarily interested in the evaluation of a light--like Wilson loop in pure $U(N)$ Chern--Simons 
theory \footnote{We refer to \cite{HPW} for notations, conventions and Feynman rules.}
\beq
\label{WLCS}
\langle W_4 \rangle_{CS} = \frac{1}{N} \left\langle {\rm Tr} \, {\cal P} \exp{i \oint_C A_\mu dx^\mu} \right\rangle
\eeq
and of the analogous object for $U(N) \times U(N)$ ABJM theory 
\beq
\label{WLABJM}
\langle W_4 \rangle_{ABJM} = \frac{1}{2N} \left\langle {\rm Tr} \, {\cal P} \exp{i \oint_C A_\mu dx^\mu} +  \hat{\rm Tr} \, {\cal P} \exp{i \oint_C \hat{A}_\mu dx^\mu} \right\rangle
\eeq
where $C$ is a tetragon with edges $p_i^\mu \equiv x_{i+1}^\mu - x_i^\mu$, $i=1, \cdots, 4$ satisfying $p_i^2 =0$.  

Perturbative evaluation suffers from short distance divergences arising near the cusps. A widely used method for regularizing integrals is based on analytical continuation of the space--time dimensions. 
In supersymmetric theories the most convenient scheme is dimensional regularization with dimensional reduction \cite{Siegel}, where 
Feynman rules are given in integer dimensions $n$, the spinorial and tensorial algebras involving objects like $\gamma$ matrices and Levi--Civita $\varepsilon$ tensors are performed strictly in $n$ dimensions, whereas momentum integrals are continued to complex $d=n-2\epsilon$.
This scheme has been applied to three--dimensional Chern--Simons theories with and without matter and has been proved to be consistent with the gauge invariance and supersymmetry of the theory \cite{CSW}. Recently, this prescription has been also shown to reproduce the results coming from localization for the case of 1/2 BPS circular Wilson loops in ABJM theory \cite{BGLP}. 

Care has to be taken when contracting objects of different dimensionality, specifically three--dimensional objects coming from Feynman rules 
with $d$--dimensional tensors arising from tensorial integrals.
DRED scheme assigns the following rules  \cite{Siegel:1980qs} for contracting three--dimensional metrics $\eta^{\mu\nu}$ (we consider Lorentzian signature $\eta^{\mu\nu} = {\rm diag}(1,-1,-1)$) and $d$--dimensional ones $\hat \eta^{\mu\nu}$ 
\begin{equation}\label{eq:DRED}
\eta^{\mu\nu} \eta_{\mu\nu} =3 \qquad \quad 
\hat{\eta}^{\mu\nu} \hat{\eta}_{\mu\nu} = 3-2\e \qquad \quad 
\eta^{\mu\nu} \hat{\eta}_{\nu\rho} = \hat \eta^{\mu}_{\phantom{\mu}\rho}
\end{equation}
Analogously, vectors $p_i^\mu$ coming from integrals have to be thought as being $d$--dimensional vectors, so that unambiguous rules are applied for contractions $p_i^\mu p_{j \, \mu}$ and $p_i^\mu \hat{\eta}_{\mu \nu}$ or $p_i^\mu \eta_{\mu \nu}$.

A concern may arise, instead, when contracting $d$--dimensional metric tensors with Levi--Civita tensors which cannot be defined outside three dimensions.  Usually, two possible strategies for overcoming the problem can be used: Either tensor algebra is performed until one reaches a situation where only scalar integrals survive \cite{CSW}, or one applies algebraic identities in order to get rid of all $\varepsilon$ tensors. 

In the following, we will adopt the second strategy to compute tetragonal Wilson loops (\ref{WLCS}, \ref{WLABJM}).

\section{The computation}

As discussed in \cite{HPW, BLMPRS}, the expressions (\ref{WLCS}, \ref{WLABJM}) vanish at one--loop. Therefore, the first non--trivial
contribution appears at two loops. 

In the planar limit, the only non--vanishing diagrams for the two--loop Wilson loop in pure Chern--Simons theory are associated to ladder and three--vertex  topologies given in Fig. \ref{fig:vertex}.

\FIGURE{ 
    \centering
    \includegraphics[width=.5\textwidth]{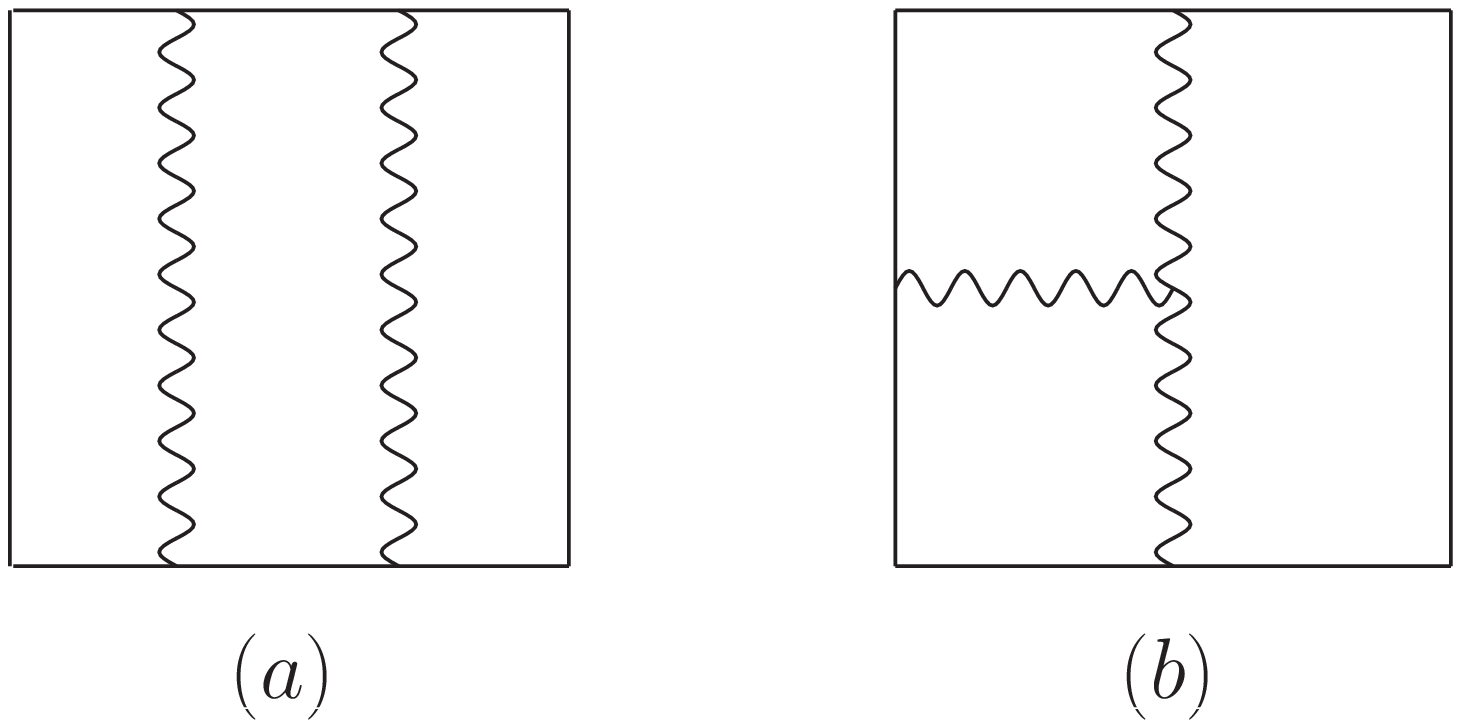}
    \caption{Gauge contributions.}
    \label{fig:vertex}}

While the ladder diagrams are finite and can be computed directly in three dimensions, divergent integrals appear from diagrams of the type $1(b)$, which have then to be dimensionally regularized. Consequently, they suffer from regularization ambiguities due to the explicit appearance of $\varepsilon$ tensors coming from the cubic vertex and the gauge propagators.

After some algebra, the contributions from $1(b)$--type diagrams can be expressed in terms of the following integral \cite{HPW}
\begin{align}
I_{321} =& \int\, d^3s_{1,2,3}\, \varepsilon^{\lambda\mu\nu}\, \varepsilon_{\rho\sigma\tau}\,
p_{1\, \lambda}\, p_{2\, \mu}\, p_{3}^{\rho}\, p_{2}^{\sigma}\, \partial_{z_1 \, \nu}\, \partial_{z_3}^{\tau} \nonumber\\&
\int\, d^d w\, \frac{(d-2)^{-2}}{(w^2)^{3/2-\epsilon} \left[(w-z_{12})^2\right]^{1/2-\epsilon} \left[(w-z_{32})^2\right]^{1/2-\epsilon}}
\end{align}
where $z_i$ indicate the position on the edge $p_i$, parameterized by $s_i$, $z_i^{\mu}=x_i^{\mu}+p_i^{\mu}s_i$. 
The method adopted in \cite{HPW} consists in first solving the scalar integral by Feynman parameterization in $d$ dimensions  
\begin{equation}
\label{integral}
I_{321}= i\, \pi^{d/2}\, \frac{\Gamma\left(d-2\right)}{\Gamma\left(3d/2-2\right)}\frac{1}{(d-2)^2} \int\, [d\beta]_3\, d^3 s_{1,2,3}\, 
\varepsilon^{\lambda\mu\nu}\, \varepsilon_{\rho\sigma\tau}\,
p_{1\, \lambda}\, p_{2\, \mu}\, p_{3}^{\rho}\, p_{2}^{\sigma}\, \partial_{z_1 \, \nu}\, \partial_{z_3}^{\tau}
\frac{1}{\Delta^{d-2}}
\end{equation}
where ($z_{ij}^\mu \equiv z_i^\mu - z_j^\mu$, $\bar \beta_i = 1-\beta_i$)
\begin{equation}
\Delta= 2\, \beta_1\,\beta_3\, (z_{12}\cdot z_{32})- z_{12}^2\, \beta_1\, \bar{\beta}_1- z_{32}^2\, \beta_3\, \bar{\beta}_3
\end{equation}
\begin{equation}
\int [d\beta]_3 \equiv \int_0^1 d\beta_1\, d\beta_2\, d\beta_3\,  (\beta_1\beta_2\beta_3)^{(d-2)/2-1}\, \beta_2\,  \delta\left(\sum_i \beta_i-1\right)\,  \frac{\Gamma(\frac{3d}{2}-2)}{\Gamma(\frac{d}{2})\Gamma(\frac{d}{2}-1)^2}
\end{equation}
and then applying the derivatives. It is important to note that, since this is equivalent to performing the derivatives first and then compute a tensor integral, the derivatives are to be considered as living in $d =(3-2\epsilon)$ dimensions.  
It follows that the application of derivatives yields
\bea
\label{I321}
&& \partial_{z_1 \, \nu}\, \partial_{z_3}^{\tau}
\frac{1}{\Delta^{d-2}} =
\\
&& -(d-2) \left[ \frac{2\, \beta_1\, \beta_3\, \hat \eta_{\nu}^{\phantom{\nu}\tau} }{\D^{d-1}} - 4(d-1)\, \frac{(\beta_1\, \beta_3\, z_{32 \, \nu} - \b_1\, \bar \b_1\, z_{12 \, \nu}) (\beta_1\, \beta_3\, z_{12}^{\tau} - \beta_3\, \bar \beta_3\, z_{32}^{\tau})}{\D^d} \right] \non
\eea
where the metric appearing in the first term is a $d$--dimensional metric.

Inserting back into $I_{321}$, the second term in (\ref{I321}), being proportional to external vectors, can be safely 
contracted with the product of $\varepsilon$ tensors as done in \cite{HPW}, giving the second piece in formula 
(\ref{eq:I123}) below. In the first term, instead, we have to evaluate
\begin{equation}
\varepsilon^{\lambda\mu\nu}\, \varepsilon_{\rho\sigma\tau}\, \hat \eta_{\nu}^{\phantom{\nu}\tau}\, p_{1\, \lambda}\, 
p_{2\, \mu}\, p_{3}^{\rho}\, p_{2}^{\sigma}
\end{equation}
where contractions of a $d$--dimensional metric with three--dimensional Levi--Civita tensors appear. To overcome the 
problem, we can get rid of $\varepsilon$ tensors by using the identity
\begin{equation}
\varepsilon^{\lambda\mu\nu} \varepsilon_{\rho\sigma\tau} =  \delta^{\lambda\mu\nu}_{[\rho\sigma\tau]} 
\end{equation}
When applied to our case, all terms containing $\eta^{\mu}_{\phantom{\mu}\sigma}$ vanish because of the light--cone 
condition, $p_2^2 = 0$. We are then left with
\begin{equation}
-\left(\hat \eta^{\rho}_{\phantom{\rho}\rho} -2 \right)\, \eta^{\lambda}_{\phantom{\lambda}\sigma}\, 
\eta^{\mu}_{\phantom{\mu}\rho}\, p_{1\, \lambda}\, p_{2\, \mu}\, p_{3}^{\rho}\, p_{2}^{\sigma}
\end{equation}
This can be evaluated using the DRED rules (\ref{eq:DRED}) and gives ($2\, p_1\cdot p_2 \equiv s$, $2\, p_2\cdot p_3 
\equiv t$)
\begin{equation}
-\frac{d-2}{4}\, s\, t
\end{equation}
Therefore, the final integral to be computed reads
\begin{align}\label{eq:I123}
I_{321}&= i\, \pi^{d/2}\, \frac{\Gamma(d-1)}{8\, \Gamma^{3}(d/2)}\, s   t \,  
\int_0^1\, d^3\, s_{1,2,3}\, d^3\,\beta_{1,2,3}\, (\beta_1\,\beta_2\,\beta_3)^{(d-2)/2}
\nonumber\\&
\delta\left(\sum_i \beta_i-1\right) \left( \frac{1}{\Delta^{d-1}}(d-2) - 2\frac{(d-1)}{\Delta^{d}}\beta_1\,\beta_3\, \bar{s}_1\, s_3\, (s+t) ) \right)
\end{align}
\FIGURE{ 
    \centering
    \includegraphics[width=.5\textwidth]{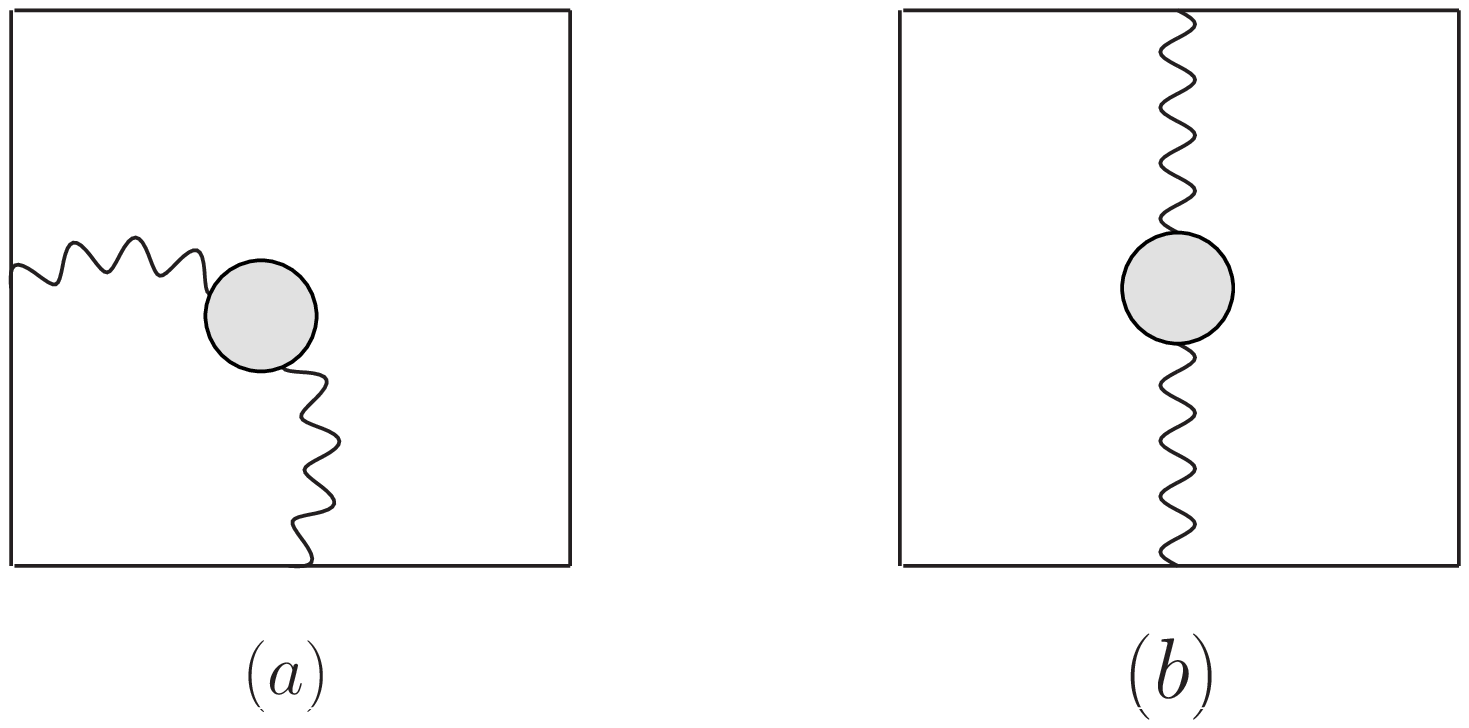}
    \caption{Matter contributions.}
    \label{fig:matter}}
We see that this expression differs from the result (B.7) in Ref. \cite{HPW} by a factor $(d-2) = (1-2\epsilon)$ in the 
first term.
This discrepancy can be traced back  to the application of DRED rules when dealing with 
$\varepsilon$ tensors. 

As discussed in \cite{HPW}, the ABJM Wilson loop (\ref{WLABJM}) is obtained by adding to the contributions from 
diagrams in Fig. \ref{fig:vertex} two extra contributions drawn in Fig. \ref{fig:matter} where one--loop matter 
corrections to the gauge propagators appear \footnote{We recall that in DRED the one--loop gauge and ghost corrections cancel exactly \cite{CSW}.}. 

While the first diagram is finite, the second one is divergent and 
requires regularization. The one--loop correction to the gauge propagator, coming only from matter contributions, contains $\varepsilon$ 
tensors. However, summing bosonic and fermionic loops, the index structure reduces to three dimensional tensors only and the $\varepsilon$ tensor algebra can be performed unambiguously, as done in \cite{HPW}. 

To summarize, both for the pure Chern--Simons and for the ABJM Wilson loops the only change which arises from a careful 
application of DRED is confined to the vertex integral (\ref{eq:I123}). In the next Section we discuss the consequences 
of this mismatch on the final result for $\langle W_4 \rangle_{\rm CS}$ and $\langle W_4 \rangle_{\rm ABJM}$.

\section{The result}

As described above, revisiting the calculation of \cite{HPW} by a careful use of DRED does not lead to any change in 
the evaluation of two--loop diagrams for planar light--like Wilson loops, except for the integral (\ref{eq:I123}) 
associated to the vertex diagram in Fig. \ref{fig:vertex}. Since the integral $I_{321}$ is $1/\e$ divergent, the 
evanescent term carried by the extra factor $(d-2)$ in (\ref{eq:I123}) modifies non--trivially the constant part of the 
two--loop result, as we now explain.  

The final result for the pure Chern--Simons Wilson loop found in \cite{HPW} by summing ladder and vertex contributions 
reads
\beq
\label{WLold}
\langle W_4 \rangle_{CS}^{(2)} = - \left( \frac{N}{k} \right)^2 \frac14 \left[ \log{(2)} \sum_{i=1}^4 
\frac{(-x_{i,i+2}^2\, \tilde \mu^2 )^{2\epsilon}}{\epsilon}
- 10 \zeta_2 + 8\log{(2)} + 8 \log^2{(2)} \right] 
\eeq
where the regularization scale has been redefined as  $\tilde \mu^2 = \mu^2 \pi e^{\gamma_E}$. This result exhibits a 
non--maximally transcendental constant $8\log{(2)}$. This leads to the speculation that the heuristic maximal transcendentality principle
\cite{Kotikov:2001sc} might not be working for light--like Wilson loops in Chern--Simons theory.   

However, if we now redo the calculation by taking into account the extra factor $(d-2)$ in eq. (\ref{eq:I123}) it is 
easy to realize that a further constant contribution $- 8 \log 2$ is produced, which exactly cancels the 
transcendentality-one term in (\ref{WLold}). 

Therefore, using the correct prescription for dealing with Levi--Civita tensors in DRED, the final expression for the 
tetragonal Wilson loop in pure Chern--Simons turns out to be
\begin{align}\label{eq:CSWL}
\langle W_4 \rangle_{\rm CS} &= 1 -\left(\frac{N}{k}\right)^2  \frac{1}{4}\left[\log(2) \sum_{i=1}^4\frac{(-x_{i,i+2}^2 \, \tilde \mu^2 )^{2\epsilon}}{\epsilon} - 10 \zeta_2 + 8 \log^2 (2) + {\cal O}(\e) \right] + {\cal O}(k^{-3})
\end{align}
and is manifestly maximally transcendental. 

The same effect occurs in the evaluation of the ABJM Wilson loop. In fact, summing to the previous result the extra contributions from diagrams in Fig. \ref{fig:matter} and redefining the mass scale as ${\mu^\prime}^2= 8 \pi e^{\gamma_E}\mu^2 $ in order to avoid the appearance of $1/\e$ poles, we find
\begin{align}
\label{WLABJM2} 
&\langle W_4 \rangle_{\rm ABJM} =1 +\frac{1}{4}\left( \frac{N}{k}\right)^2 \left[ - \frac{(- 
{\mu^\prime}^2 \, x_{13}^2)^{2\epsilon}}{\epsilon^2} -
\frac{(- {\mu^\prime}^2 \, x_{24}^2)^{2\epsilon}}{\epsilon^2}  + 2  \log^2\left(\frac{x_{13}^2}{x_{24}^2}\right) \right.
\nonumber\\& \hspace{6cm} + 16 \zeta_2 + 12 \log^2(2) + {\cal O}(\e)  \Big] + {\cal O}(k^{-3})
\end{align}
We observe that it does have maximal transcendentality, exactly like its analogue in ${\cal N}=4$ SYM 
\cite{Drummond:2007aua, Brandhuber:2007yx, Drummond:2007cf} 
and its dual object, the four--point scattering amplitude \cite{CH, BLMPS1}. In particular, expressing the four--point 
amplitude in terms of dual variables $p_i=x_{i+1}-x_i$ and identifying them with the Wilson loop coordinates, the result  
(\ref{WLABJM2}) not only matches the divergent and $\log^2\left(\frac{x_{13}^2}{x_{24}^2}\right)$ parts of the 
scattering amplitude, but it also matches its maximally transcendental constant.

\vskip 10pt

The result we have found opens the possibility that light--like Wilson loops in ABJM theory could enjoy maximal 
transcendentality, as it appears to be the case in ${\cal N}=4$ SYM. To test this conjecture, the educated use of DRED 
that we have applied in the evaluation of the tetragonal Wilson loop should be extended 
to $n$ cusped Wilson loops \cite{Wiegandt}, with effects on the constant part of the result. However, for generic $n$ 
the analysis of transcendentality is hampered by the lack of analytical results for the constants.

The correctness of our prescription for dealing with Levi--Civita tensors in DRED is proved by the fact that it has 
played a crucial role in the recent computation of $1/2$ BPS circular Wilson loops, in particular in matching the 
perturbative result with the prediction from localization \cite{BGLP}.

\section*{Acknowledgements}
We thank Johannes Henn, Jan Plefka, Konstantin Wiegandt and Gabriele Travaglini for interesting comments.
MB thanks L. Bianchi and V. Forini for useful discussions. 
The work of MB has been supported by the Volkswagen-Foundation.
The work of GG and ML has been supported by the research project CONICET PIP0396.
The work of SP has been supported in part by INFN, MIUR--PRIN contract 2009--KHZKRX and MPNS--COST Action MP1210 "The 
String Theory Universe".

\end{document}